# Detection of a Cosmic Ray with Measured Energy Well Beyond the Expected Spectral Cutoff Due to Cosmic Microwave Radiation

D.J. Bird,[1,3] S.C. Corbató,[2] H.Y. Dai,[3] J.W. Elbert,[3] K.D. Green,[4] M.A. Huang,[3] D.B. Kieda,[3] S. Ko,[3]
C.G. Larsen,[3] E.C. Loh,[3]
M.Z. Luo,[5] M.H. Salamon,[3] J.D. Smith,[3] P. Sokolsky,[3] P. Sommers,[3] J.K.K. Tang,[3] S.B. Thomas,[3]

[1]*Department of Physics, University of Illinois at Urbana-Champaign, Urbana, IL 61801 USA*
[2]*Present address: NorthWestNet, Bellevue, WA 98007*
[3]*High Energy Astrophysics Institute,*
*Department of Physics, University of Utah, Salt Lake City UT 84112 USA*
[4]*Present address: Department of Physics, The University of Michigan, Ann Arbor, MI 48109 USA*
[5]*Present address: Biology Division, Dugway Proving Ground, Dugway, UT 84022*

## Abstract

We report the detection of a 51-joule ($3.2 \pm 0.9 \times 10^{20}$ eV) cosmic ray by the Fly's Eye air shower detector in Utah. This is substantially greater than the energy of any previously reported cosmic ray. A Greisen-Zatsepin-Kuz'min cutoff of the energy spectrum (due to pion photoproduction energy losses) should occur below this energy unless the highest energy cosmic rays have traveled less than about 30 Mpc. The error box for the arrival direction in galactic coordinates is centered on $b = 9.6°$, $l = 163.4°$. The particle cascade reached a maximum size near a depth of 815 g/cm$^2$ in the atmosphere, a depth which does not uniquely identify the type of primary particle.

## 1 Introduction

The existence of cosmic rays with energies above 100 EeV ($100 \times 10^{18}$ eV) is of special interest because particles of such high energy cannot propagate freely through the cosmic background radiation. In the restframe of such an energetic proton, the cosmic microwave radiation constitutes a beam of gamma rays, many of which are energetic enough to collide with the proton and produce a pion. In the universal restframe (in which the microwave radiation is at rest), the energetic proton is seen to lose energy as a result of the pion photoproduction. If the sources of the highest energy cosmic rays were all at cosmological distances, the energy spectrum would exhibit a Greisen-Zatsepin-Kuz'min (GZK) cutoff below 100 EeV (Greisen, 1966; Zatsepin & Kuz'min, 1966). If the sources are at a distance of approximately 100 Mpc, the expected spectral cutoff is less sharp, but the cosmic ray intensity is markedly attenuated at energies above 100 EeV (Stecker, 1968; Hill & Schramm, 1985; Yoshida & Teshima, 1993). For protons above 300 EeV, the attenuation length is less than 30 Mpc (Stecker, 1968). Nuclei and gamma rays have even shorter survival times (Puget et al., 1976; Wdowczyk et al., 1972).

It is therefore significant that the Fly's Eye detector in Utah recorded a cosmic ray air shower whose energy was approximately 320 EeV. The source of this particle should be sought within about 30 Mpc. Due to its high magnetic rigidity, its arrival direction may point approximately toward its point of origin. The arrival direction in 1950 celestial coordinates is $\alpha = 85.2°\pm 0.4°$ and $\delta = 48.0°^{+5.2°}_{-6.3°}$. Its detection in Universal Time occurred at 7:34:16 on October 15, 1991. This is the only Fly's Eye air shower with energy greater than 80 EeV.

An uncertainty of 93 EeV is associated with the energy of this shower. This includes both systematic and statistical uncertainties added in quadrature (cf. Table 1). The possible systematic error comes from uncertainty in the atmospheric scintillation efficiency, uncertainty in light attenuation due to the variable atmospheric aerosol concentration, and approximations implicit in the way we parametrize air showers. The statistical uncertainty emerges from a specific least squares fitting procedure which is used to determine the shower parameters from recorded data. As part of this procedure, a detailed model of the detector is used to derive the expected data for any trial set of shower parameters. Our method for reconstructing the air shower from the recorded data is summarized below. The statistical uncertainty in the energy is dominated by possible error in the fitted value for the distance to the shower axis, which is determined by a non-linear fit to phototube trigger times. An energy lower bound which is independent of that detailed fit can be obtained by assuming that the first interaction did not occur anomalously deep in the atmosphere. That method also shows this particle's energy to be beyond the GZK cutoff.

Table 1: *Shower Parameters.* Statistical uncertainties are derived using the same $\chi^2$ function whose minimization defines the best-fit values. This $\chi^2$ function depends on our modeling of the light production, atmospheric transmission, and detector response. The systematic uncertainties include contributions from possible errors in each of those models. The last column of the table has been obtained by adding the statistical and systematic uncertainties in quadrature. Uncertainties in galactic coordinates are not shown in order to emphasize that the error box happens to have a simple rectangular form in declination and right ascension.

| | Best Fit Value | Statistical Uncertainty | Systematic Uncertainty | Combined Uncertainty |
|---|---|---|---|---|
| Energy | 320 $EeV$ | $^{+35\ EeV}_{-40\ EeV}$ | $\pm 85\ EeV$ | $^{+92\ EeV}_{-94\ EeV}$ |
| $X_{max}$ | 815 $g/cm^2$ | $^{+45\ g/cm^2}_{-35\ g/cm^2}$ | $\pm 40\ g/cm^2$ | $^{+60\ g/cm^2}_{-53\ g/cm^2}$ |
| $R_p$ | 13.0 $km$ | $^{+0.5\ km}_{-0.8\ km}$ | $\pm 0.8\ km$ | $^{+0.9\ km}_{-1.1\ km}$ |
| $\psi$ | 76.6° | $^{+3.2°}_{-4.8°}$ | $\pm 4.1°$ | $^{+5.2°}_{-6.3°}$ |
| *Right Ascension* | 85.2° | $\pm 0.2°$ | $\pm 0.5°$ | $\pm 0.5°$ |
| *Declination* | 48.0° | $^{+3.2°}_{-4.8°}$ | $\pm 4.1°$ | $^{+5.2°}_{-6.3°}$ |
| *Galactic Latitude* | 9.6° | – | – | – |
| *Galactic Longitude* | 163.4° | – | – | – |
| $\theta$ (*zenith angle*) | 43.9° | $^{+1.4°}_{-0.6°}$ | $\pm 1.2°$ | $^{+1.8°}_{-1.3°}$ |
| *Plane Normal Dec* | −0.63° | $\pm 0.15°$ | $\pm 0.5°$ | $\pm 0.5°$ |
| *Plane Normal R.A.* | −4.33° | $\pm 0.05°$ | $\pm 0.2°$ | $\pm 0.2°$ |

Other experiments have produced evidence for air showers with energies above 100 EeV (Linsley 1963; World Data Center for Cosmic Rays, 1980 & 1986; Efimov et al., 1991). Due to statistical uncertainties and possible systematic experimental errors in energy determinations, those showers are not unambiguously beyond the GZK cutoff. The measured energy of the shower reported here is more than twice as great as any of the previously reported energies and is well beyond the GZK energy cutoff expected for distant sources. It constitutes strong evidence for a nearby source of superhigh energy cosmic rays.

## 2  The shower data and analysis

The Fly's Eye is a compound eye of 880 photomultiplier tube (PMT) pixels which collectively monitor the $2\pi$ steradians of visible sky (Baltrusaitis et al., 1985). The phototubes are arranged in clusters at the focal planes of 67 different mirrors, each mirror having a diameter of 1.5 meters. The detector operates on clear moonless nights. The Fly's Eye observes an air shower as a fluorescent light source which moves at the speed of light down a line through the atmosphere (the shower axis). The intensity of the light source is proportional to the number of charged particles, so the measured light intensity from different atmospheric depths along the shower axis can be converted to a "longitudinal profile," which is the number of shower particles (shower "size") as a function of atmospheric depth measured in $g/cm^2$. The integral of that longitudinal profile yields the energy of the incident cosmic ray (Sokolsky et al., 1992).

At a second site 3.4 km away, there is a partial eye (Fly's Eye II) which monitors the half of the visible sky toward the first site. This superhigh energy air shower landed in the blind side of Fly's Eye II, so it was not seen stereoscopically. The analysis of the superhigh energy shower, like the majority of Fly's Eye events, is based on monocular data.

Figure 1 indicates the pointing directions of the 22 phototubes which triggered in connection with this air shower. They are plotted on the hemisphere of monitored directions, and each PMT has a field of view which is 5.5° in diameter. For each of these triggered tubes, the data include its trigger time (when the light flux to that tube first exceeded a threshold value) and its amplitude (the time-integral of the light flux as the shower front moved through its field of view). There is a well-determined plane which contains the Fly's Eye location and which most nearly accommodates the directions of all the triggered phototubes. The orientation of this "track-detector plane" is characterized by its unit normal vector, whose celestial coordinates are given in Table 1. The shower axis necessarily lies in this track-detector plane. By chance, the plane for this shower's track nearly contains the celestial poles, so the right ascension of the shower arrival direction is well determined by the track-detector plane by itself, whereas its declination depends on the shower axis within the plane. Uncertainty in the determination of the plane is dominated by uncertainties in the pointing directions of phototube clusters, which are accurate to 0.3°.

Identification of the particular line in the track-detector plane which corresponds to the shower axis must be accomplished using the PMT trigger times. There is a 2-parameter family of possible shower axes in the plane. The two parameters may be chosen as $R_p$ and $\psi$, as indicated in Figure 2. $R_p$ is the perpendicular distance from the Fly's Eye to the axis, and $\psi$ is the angle the axis makes with the horizontal line that lies in the track-detector plane. The PMT trigger times give a mean angular speed for the shower, as viewed by the Fly's Eye. Since the shower front is known to move at the speed of light, the mean angular speed along the detected track gives $R_p$ as a function of $\psi$ or $\psi$ as a function of $R_p$. The mean angular speed therefore reduces the set of possible shower axes to a 1-parameter family of lines in the plane. The remaining independent parameter may be chosen as either $R_p$ or $\psi$. The difficulty in picking out the correct axis from this 1-parameter family of lines is responsible for most of the uncertainty in the shower's direction and its longitudinal profile. The true shower axis must be identified by finding the line which yields not only the observed mean angular speed but also the small deviations from constant angular speed.

The method is, in effect, to try every line in the track-detector plane as a hypothesis for the shower axis, and find the one for which the expected PMT trigger times agree best with the actual trigger times. The expected trigger times depend not only on the shower axis, but also on the amplitude of the signal in each tube. This is partly due to the response of an electronic filter and partly due to spreading of the shower image in the focal plane by optical aberrations and distortion in the mirrors. In testing each trial shower axis, we have therefore used the appropriate amplitude for each tube and simulated the optics and electronics in order to calculate the expected trigger time for each PMT. Mirror distortions are not modeled, however, and they are the major source of unpredictable jitter in trigger times for large amplitude tubes. (The distortion may produce a systematic effect in causing the bright tubes late in the shower development to trigger early, while having little effect on the low-amplitude tubes near the start of the shower. Incorporating that systematic effect would raise the energy estimate slightly above 320 EeV.) Photoelectron statistical fluctuations are important only in the low-amplitude tubes. The best fit shower axis is indicated explicitly in Table 1 by the values of $R_p$ and $\psi$ and implicitly in the different shower direction coordinates.

For a known shower axis, the longitudinal profile (and energy) can be determined from the amplitudes of phototubes which view different parts of the shower axis. Light production is modeled carefully, including direct Cherenkov light and light scattered to the Fly's Eye from the intense Cherenkov beam along the shower axis. For this shower, the direct Cherenkov light is negligible because the line of sight to every part of the detected shower axis makes an angle greater than 40° with the shower axis. The scattered Cherenkov beam makes only a small contribution to the total detected light flux. It accounts for less than 1% of the flux when the shower is first seen, increasing to 28% at ground level. Simulation of the light propagation from the shower axis to the Fly's Eye includes Rayleigh scattering, aerosol scattering, and ozone absorption. After subtracting the expected scattered Cherenkov light and correcting for atmospheric attenuation, the light flux at the detector can be used to determine the fluorescent light intensity (and hence the number of shower particles) in intervals along the shower axis. We use three parameters to characterize a longitudinal profile: $S_{max}$ is the maximum size attained; $X_{max}$ is the atmospheric depth where the size $S_{max}$ occurs; and $X_0$ is the depth at which the cascade originates. These parameters define a longitudinal profile by the Gaisser-Hillas functional form (Gaisser & Hillas, 1977):

$$Size(X) = S_{max} \cdot \left(\frac{X - X_0}{X_{max} - X_0}\right)^{(X_{max} - X_0)/70} exp[(X_{max} - X)/70],$$

where $X$ is atmospheric depth in g/cm². Figure 3 shows the best-fit Gaisser-Hillas profile along with data binned in 5°-intervals along the shower axis. The energy and $X_{max}$ for this profile are recorded in Table 1.

All together, there are 7 parameters to be fitted: $R_p, \psi, X_0, X_{max}, S_{max}$, and two independent components for the unit vector normal to the track-detector plane. The PMT amplitudes are used in determining the plane's normal vector and the three shower profile parameters. The trigger times are used in fitting $R_p$ and $\psi$. Although these two $\chi^2$ minimization procedures fit different data in order to determine different shower parameters, they are not completely independent. The choice of shower axis in the track-detector plane affects the best-fit shower profile, and the plane and shower profile parameters affect the expected PMT trigger times. We therefore use an iterative procedure in which the amplitude and timing chisquares are alternately minimized. (These reconstruction procedures are described in detail elsewhere (Dai, 1994).) There are 14 phototubes whose directions are close enough to the track-detector plane for the simulation program to compute expected data for them. The other tubes are low amplitude tubes which triggered because of scattered light and mirror distortions. The amplitudes and trigger times for the 14 phototubes are 28 data for determining the 7 parameters. For the best-fit solution given in Table 1, the amplitude $\chi^2$ is 1.1 per degree of freedom. The timing $\chi^2$ does not have a meaningful absolute value. The relative PMT trigger time uncertainties are computed by simulation, but the normalization of the uncertainties is not fixed because mirror distortions have not been modeled. The $\chi^2$ minimization is independent of that normalization. The normalization for the PMT timing jitter is here chosen so that the sum of the two minimized chisquares is exactly 21 (i.e. 1 per degree of freedom).

Except for the unit normal vector to the track-detector plane, all of the shower parameters have statistical errors which are dominated by the difficulty in resolving the shower axis within the track-detector plane. Figure 4 displays the uncertainties in the two parameters $R_p$ and $\psi$ which characterize the shower axis in that plane. The elongated contour lines demonstrate that $R_p$ and $\psi$ are highly correlated, so the possible shower axes can be effectively labeled by a single parameter. Figure 5 displays the 1-parameter dependence of the $\chi^2$ function. The independent parameter may be taken to be the shower impact parameter $R_p$. With any fixed value for $R_p$, the combined $\chi^2$ can be minimized by adjusting the other 6 parameters. Each fixed $R_p$ gives a $\chi^2$ value, as shown in the figure. It also yields a best-fit energy, $X_{max}$, and declination. The figure shows the $\chi^2$ also as a function of those correlated variables. The plots can be used to estimate the uncertainties in those quantities arising from the least squares fitting, by noting the range over which the $\chi^2$ function is less than 1 unit above its minimum value.

Figure 6 displays the trigger time and amplitude for each of the 14 phototubes. Also shown in Figure 6 are the expected amplitudes and trigger times, based on numerical simulation using the best-fit shower parameters. It can be seen that the deduced shower geometry and profile lead to expected data which are in good agreement with the recorded data.

In addition to statistical uncertainties in the shower parameters arising from fluctuations in the measured quantities, some systematic errors also pertain. Both types of uncertainties are indicated in Table 1. The possible systematic errors in $X_{max}$ and energy have been discussed in detail elsewhere (Bird et al., 1994; Gaisser et al., 1993). A major systematic uncertainty for a distant event like this one is the aerosol concentration in the atmosphere at the time of detection. That is monitored locally in an approximate way by comparing the amount of light scattered at small angles vs. the amount scattered at large angles from a flasher beam which is directed from Fly's Eye II over the Fly's Eye I site. The aerosol concentration on the night of this shower was consistent with the clear atmosphere model which pertains to Dugway weather in October (Sokolsky, 1993). The energy would be overestimated only if the aerosol concentration was less than that. For the unrealistic extreme hypothesis of zero aerosols, the energy is still exceptionally high at 220 EeV.

## 3  Consistency checks

The detector was in good working order at the time of this air shower, and the weather was clear. Besides routine electronic and optical calibration procedures which are incorporated in the data processing, performance of the detector is monitored by observing light scattered from vertical light beams which are pulsed periodically from numerous sites around the Fly's Eye. The PMT responses to those flashers were stable and normal near the detection time of this shower.

As a check on some aspects of the shower reconstruction, we performed a "laser replay." A pulsed nitrogen laser was taken to the location where the reconstructed shower axis meets the ground. It was fired (upward) along the reconstructed shower axis with a variety of pulse energies. The set of triggered phototubes depended on the pulse energy, and the set of triggered low-amplitude tubes varied pulse to pulse, but the primary tubes were the same as in the actual event. Moreover, the reconstructed beam-detector plane for the laser shot events, on average, differed by only 0.2° from the known plane containing the detector and the laser beam (i.e., comparing the normal vectors to the planes). The accuracy of pointing the laser was not much better than that, so a portion of that error may be due to inaccurate pointing of the laser. This good reconstruction of the beam-detector plane for the laser beam suggests that the systematic uncertainties assigned for the shower's track-detector plane in Table 1 (due primarily to possible errors in the pointing directions of phototube clusters) might be conservative. From pulse to pulse, the laser beam reconstructions produced beam-detector planes which fluctuated by 0.13°, consistent with the statistical error listed in Table 1 for the actual shower, which was evaluated by examining how the $\chi^2$ varies with changes in the assumed plane for the actual shower. Using calculations of the light emitted to the Fly's Eye from the laser beam as the result of Rayleigh and aerosol scattering, the laser pulses provided a known light source, whose intensity could be adjusted to be comparable to the fluorescent light source at various points along the shower development. There were no amplitude anomalies in this test, which verifies that the recorded PMT amplitudes have been properly converted to the light source intensities along the reconstructed shower axis. The laser replay could not be expected to replicate the pattern of trigger times because the light pulse went upward along the shower axis instead of downward.

It is of interest to seek a lower energy bound which does not rely on the non-linear timing fit to the PMT trigger times. For any value of $R_p$, there is a unique shower axis which is compatible with the mean angular speed for the angular interval of the detected track. The PMT amplitudes then give a longitudinal shower profile (and hence an energy) for any trial $R_p$. As shown in Figure 4, energy decreases and atmospheric depth increases if $R_p$ is reduced. To decrease the energy to 100 EeV would require $X_{max}$ to have the implausibly high value of 1335 g/cm$^2$. Moreover, as shown in Figure 7, the shower profile for that hypothetical geometry ($R_p$=7.0 km) starts too deep in the atmosphere. The inferred depth of first interaction $X_0$ is 807 g/cm$^2$. The mean free path is less than 40 g/cm$^2$ at 100 EeV for protons, nuclei, or gamma rays. A first interaction as deep as 807

g/cm$^2$ would therefore require the extremely unlikely penetration to more than 20 times the mean free path. Independent of the non-linear timing fit, this argument (based on expectations for the atmospheric depth at which a particle should interact) provides strong evidence that the shower's energy is beyond the GZK spectral cutoff. (The depth of first interaction is not determined as reliably as $X_{max}$ because it is sensitive to fluctuations in the low-amplitude phototubes near the start of the shower and to the specific functional form which is fitted. The fact that Fly's Eye II did *not* trigger on the event provides a useful constraint on its value. For the shower axis, $S_{max}$, and $X_{max}$ of Figure 7, Fly's Eye II would have triggered on any Gaisser-Hillas profile with $X_0$ less than 550 g/cm$^2$.)

## 4 Discussion

Because the energy of this air shower is extraordinary, one might wonder if the Fly's Eye energy determinations could be systematically high by more than the 20% systematic uncertainty which we evaluated (Bird et al., 1994). In comparing the energy spectra of different experiments recently, Teshima (1993) noted that the spectral shapes and normalizations can be brought into agreement by minor shifts in energy scales. That comparison indicated that the Fly's Eye energy determinations may be systematically *lower* than energy estimates by ground array experiments. Although a detailed analysis of detector resolutions was not performed (so the comparison of spectra could be misleading), it does suggest that if the Fly's Eye is overestimating air shower energies, then the ground arrays are overestimating them even more.

It is also relevant to ask whether this event could have been something other than a cosmic ray air shower, even though it has no peculiar property other than its extraordinary size. Suggested alternative explanations include an ordinary meteor, a relativistic dust grain, or a strangelet (Alcock & Olinto, 1988). The Fly's Eye does not trigger on ordinary meteors because of their low angular velocity. A solar system particle enters the atmosphere with speed less than $10^{-3}c$, so it could mimic the angular speed of an air shower only if it were a thousand times closer to the detector, i.e. at a distance of some meters instead of kilometers. Because meteors almost always burn up before reaching the surface, it is not expected that the Fly's Eye would detect one at such close range. Moreover, the perpendicular distance to this shower axis is known to be greater than 100

m. This is because the different mirror units which detected the shower are separated by distances on the order of 10 m. Had the shower been closer than 100 m, there would have been evident parallax between the images from the different mirrors, but such parallax was not present. Reversing the above argument, the undetectable parallax gives a lower limit of $0.05c$ for the velocity of the light source. There is also a range of distances (1-4.5 km) which can be excluded by the fact that Fly's Eye II did not trigger. That excludes the velocity range $0.2c - 0.6c$. The only possible velocities are relativistic ($v > 0.6c$) and the range $0.05c < v < 0.2c$. In the relativistic case, the reconstruction requires the depth of maximum light emission to be at least as great as 815 g/cm$^2$ (the depth for the $v = c$ reconstruction). For the slower case, it would have to be very much deeper than that. A dust grain (relativistic or not) should flare high in the atmosphere and not penetrate so deep. More exotic forms of matter (strange quark matter) have been postulated to explain unusual events observed by balloon experiments (Price et al., 1978) and high mountain experiments (Lattes et al., 1980). We cannot rule out the possibility that the superhigh energy event was a strangelet with velocity between $0.05c$ and $0.2c$. In that case, however, it had to penetrate to a slant depth of approximately 2000 g/cm$^2$, then brighten and fade rapidly in less than 100 g/cm$^2$. Moreover, the light intensity as a function of depth in that range had to mimic the *shape* of an air shower's longitudinal development shown in Figure 3. Although it is impossible to rule out every conceivable alternative, we emphasize that this event has no property to suggest that it is anything other than a cosmic ray air shower.

The longitudinal profile of the Fly's Eye shower does not identify the primary particle type. The best-fit $X_{max}$ value is consistent with the expectation for a mid-size nucleus (Gaisser et al., 1993). However, in view of the uncertainty in $X_{max}$ and fluctuations in shower development, it could have been a nucleon or a heavy nucleus. It might even have been a gamma ray. Its arrival direction is nearly perpendicular to the local geomagnetic field. A gamma ray of such high energy would likely initiate an electromagnetic cascade in the earth's magnetosphere (McBreen & Lambert 1981) and enter the atmosphere as a superposition of lower-energy electromagnetic particles. This would cause the air shower to reach maximum size earlier than the 1050 g/cm$^2$ given by the Greisen formula (Greisen, 1965) for an electromagnetic cascade at 320 EeV. (This early development contrasts with the expectation for gamma rays which do not encounter a transverse field. Due to the LPM effect (Landau & Pomeranchuk, 1953; Migdal, 1957; Mizumoto, 1993), they should develop even deeper than a Greisen formula shower.)

A high resolution Fly's Eye ("HiRes") is being constructed in Utah (Bird et al., 1993). It will have a large aperture for detecting air showers like this one, and its measurements will provide accurate geometrical reconstructions with excellent energy and $X_{max}$ resolutions.

In the meantime, this one air shower should be a useful clue to the origins of the highest energy cosmic rays. In view of its high magnetic rigidity, its arrival direction may point approximately toward its site of origin. Because it could not have propagated freely through the cosmic background radiation, that production site should lie within a distance of about 30 Mpc.

*Acknowledgements.* We are indebted to Colonels Frank Cox and James King and the staff of the Dugway Proving Grounds for their continued cooperation and assistance. This work has been supported in part by the National Science Foundation (grant PHY–91–00221 at Utah) and the U.S. Department of Energy (grant FG02–91ER40677 at Illinois).

REFERENCES

Alcock, C. & Olinto, A. 1988, *Rev. nucl. part. Sci.* **38**, 161

Baltrusaitis, R.M. et al. 1985 *Nucl. Instr. Meth.* **A240**, 410

Bird, D.J. et al. 1993, *Proc. 23rd Int. Cosmic Ray Conf. (Calgary)* **2**, 458

Bird, D.J., et al. 1994, *Ap. J.* **424**, 491

Dai, H.Y. 1994, Utah Technical Note #940501, High Energy Astrophysics Institute, University of Utah

Efimov, N.N., et al. 1991 *Astrophysical Aspects of the Most Energetic Cosmic Rays*, Ed: Nagano, M. & Takahara, F., World Scientific, Singapore, 20

Gaisser, T.K. & Hillas, A.M. 1977 *Proc. 15th Int. Cosmic Ray Conf. (Plovdiv)* **8**, 353

Gaisser, T.K., et al. 1993 *Phys. Rev.* **D47**, 1919

Greisen, K. 1965, *Progress in Cosmic Ray Physics* Ed: Wilson, J.G., North-Holland,

Amsterdam, vol. **3**, 17

Greisen, K. 1966, *Phys. Rev. Lett.* **16**, 748

Hill, C.T. & Schramm, D.N. 1985, *Phys. Rev.* **D31**, 564

Landau, L. & Pomeranchuk, I. 1953 *Dokl. Akad. Nauk (USSR)* **92**, 535 and 735

Lattes, C.M.G., Fujimoto, Y., & Hasegawa, S. 1980, *Phys. Rep.* **65**, 151

Linsley, J. 1963, *Phys. Rev. Lett.* **10**, 146

McBreen, B. & Lambert, C.J. 1981, *Proc. 17th Int. Cosmic Ray Conf. (Paris)* **6**, 70

Migdal, A.B. 1957, *JETP (USSR)* **32**, 633

Mizumoto, Y. 1993, *Proc. Tykyo Workshop on Techniques for the Study of Extremely High Energy Cosmic Rays*, Ed: Nagano, M., 194

Price, P.B., et al. 1978, *Phys. Rev.* **D18**, 1382

Puget, J.L., Stecker, F.W., & Bredekamp, J.H. 1976, *Ap. J.* **205**, 638

Stecker, F.W. 1968, *Phys. Rev. Lett.* **21**, 1016

Sokolsky, P., Sommers, P., & Dawson, B.R. 1992, *Phys. Rep.* **217**, 225

Sokolsky, P. 1993, *Tokyo Workshop on Techniques for the Study of Extremely High Energy Cosmic Rays*, Ed: M. Nagano, Institute for Cosmic Ray Research, University of Tokyo, 280

Teshima, M. 1993, *Proc. 23rd Int. Cosmic Ray Conf. (Calgary)* **Rapporteur Volume**, 257

Wdowczyk, J., Tkaczyk, W., & Wolfendale, A.W. 1972 *J. Phys. A* **5**, 1419

World Data Center for Cosmic Rays 1980, *Catalogue of Highest Energy Cosmic Rays*, No. 1, Institute of Physical and Chemical Research, Itabashi, Tokyo

World Data Center for Cosmic Rays 1986, *Catalogue of Highest Energy Cosmic Rays*, No. 2, Institute of Physical and Chemical Research, Itabashi, Tokyo

Yoshida, S. & Teshima, M. 1993, *Prog. Theor. Phys.* **89**, 833

Zatsepin, G.T. & Kuz'min, V.A. 1966 *JETP Let.* **4**, 78

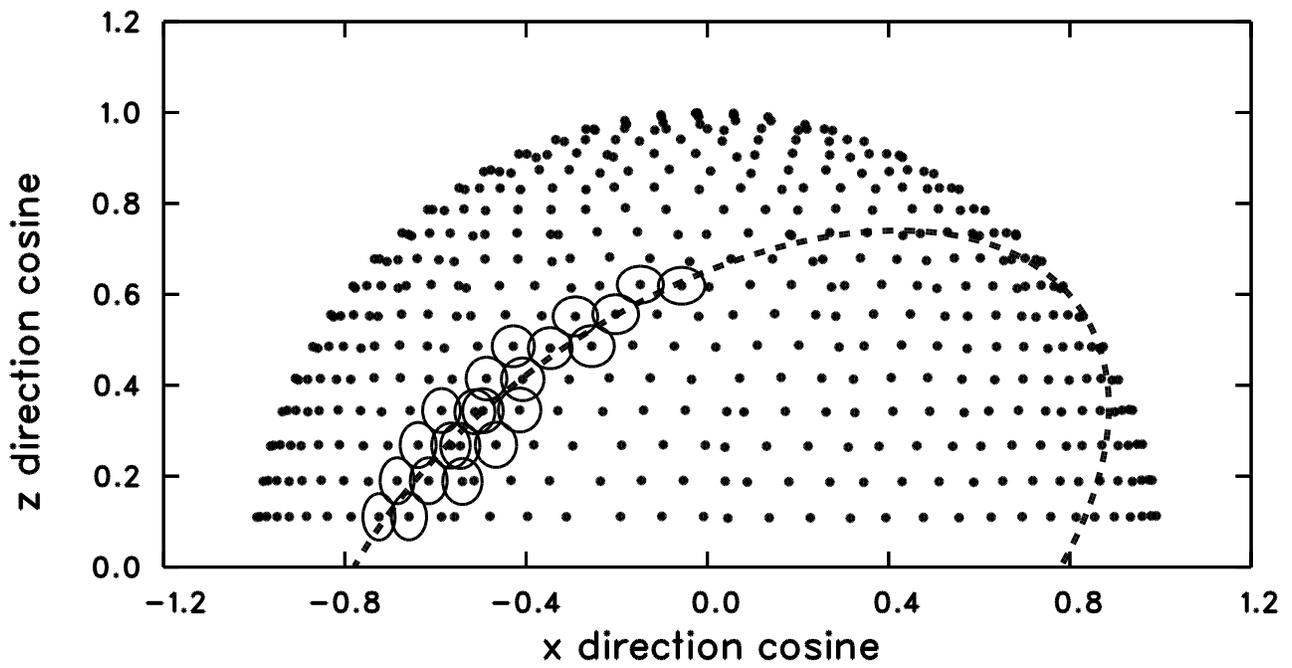

Figure 1: The pointing directions of the 22 phototubes which triggered in connection with this event are shown projected into the xz-plane. The x-axis points east, the y-axis north, and the z-axis upward. The triggered phototubes have positive y-components.

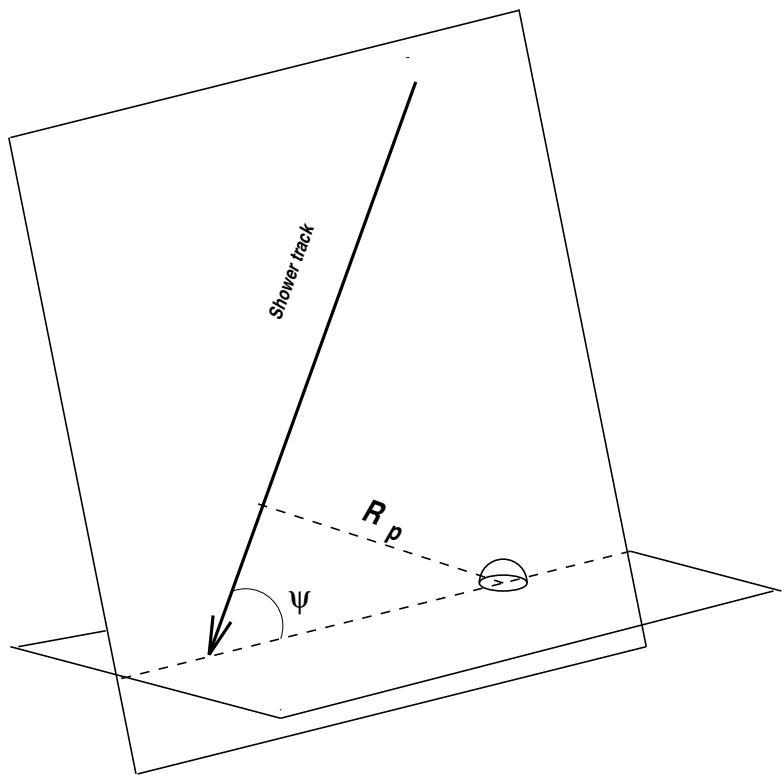

Figure 2: A shower axis in the track-detector plane is labeled by its perpendicular distance from the detector $R_p$ and the angle $\psi$ which it makes with the horizontal line in that plane.

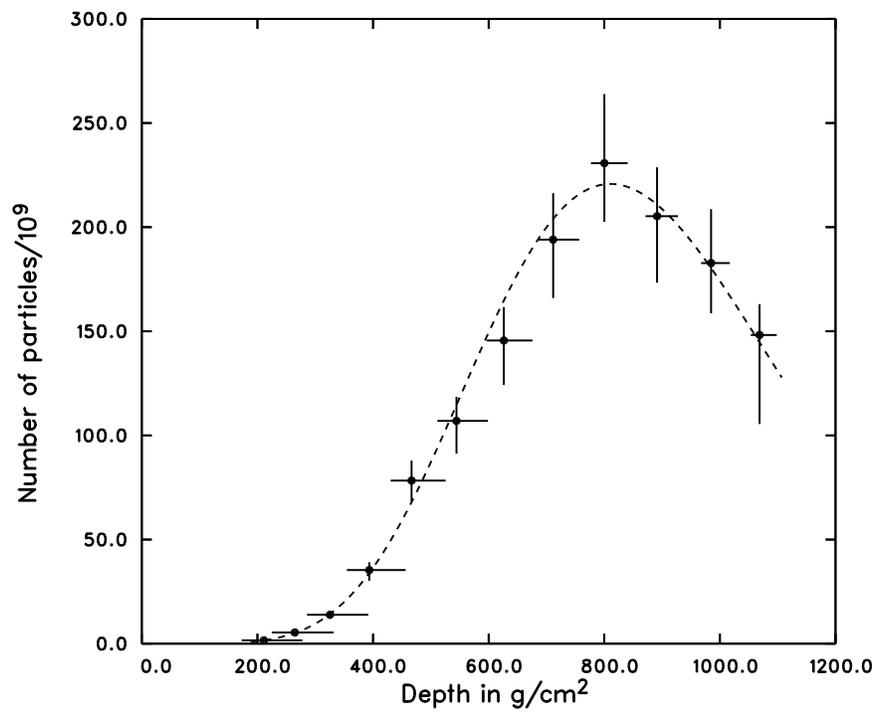

Figure 3: The 3-parameter best-fit shower profile is shown along with points obtained from the data in 5-degree intervals. The size at maximum is greater than 200 billion particles.

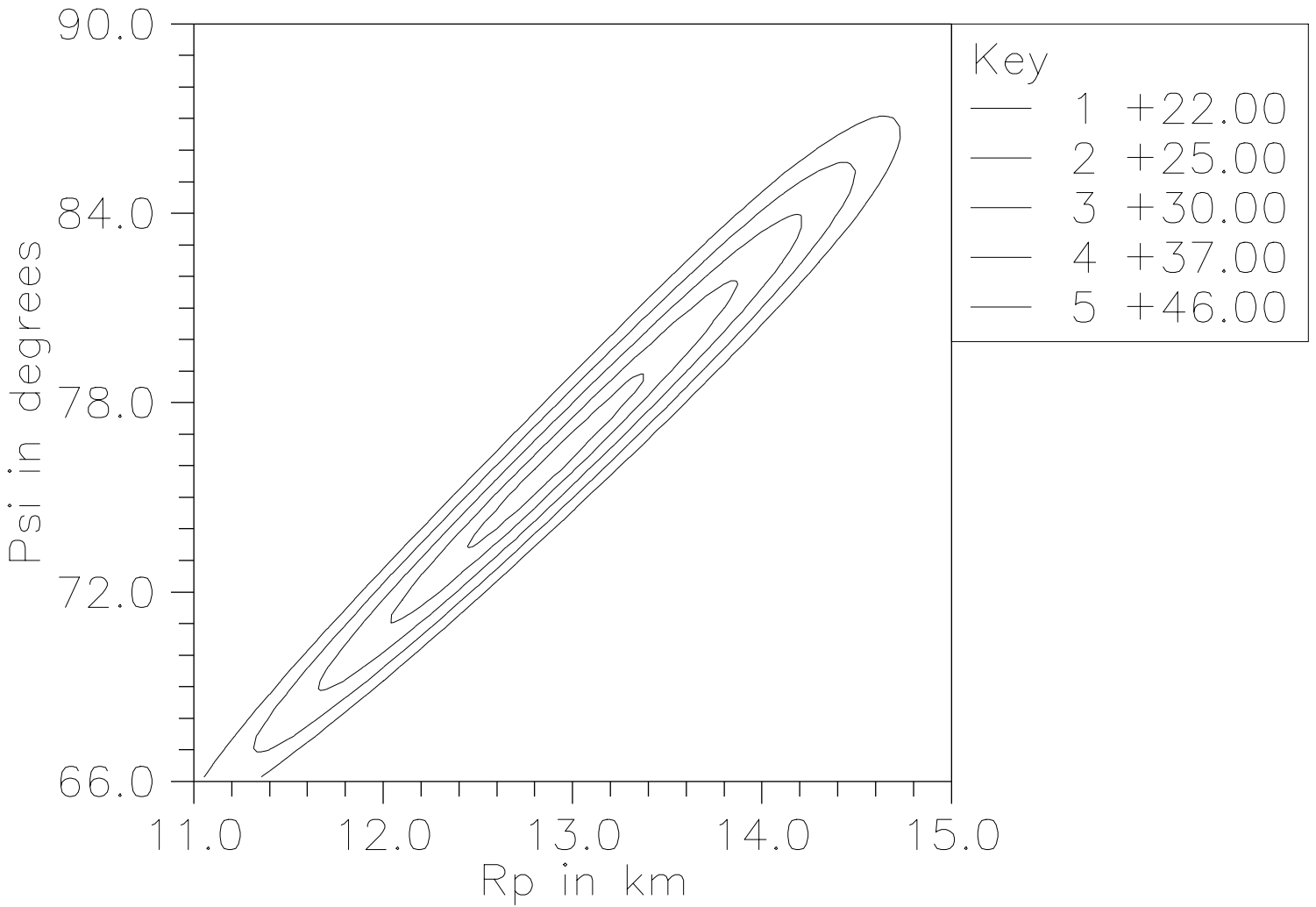

Figure 4: The $\chi^2$ function depends on the parameters which characterize a shower axis in the track-detector plane. Here $R_p$ and $\psi$ have been varied independently, and the $\chi^2$ has been minimized with respect to the other parameters. If projected onto either axis, these five contours give the $1\sigma$, $2\sigma$, ..., $5\sigma$ uncertainty ranges. The elongated contour lines illustrate that there is effectively only a single degree of freedom in the fit.

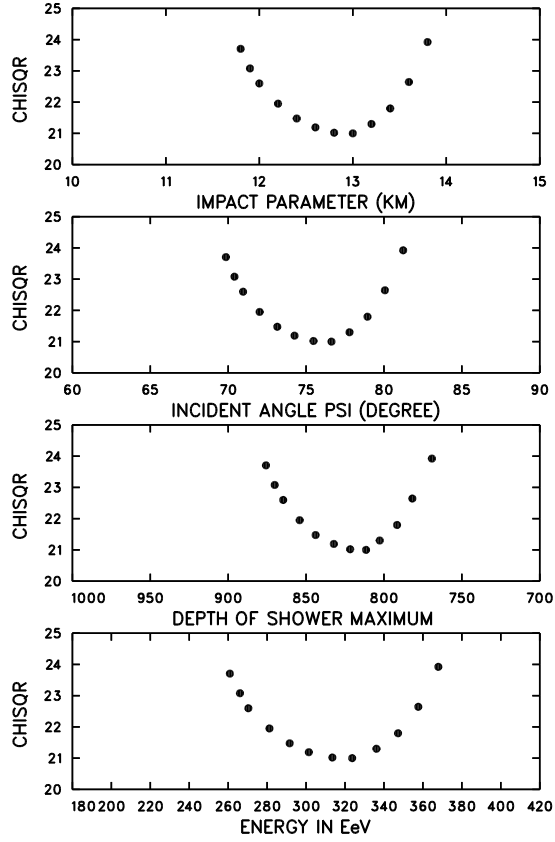

Figure 5: The statistical (fitting) uncertainties in shower parameters are dominated by a 1-parameter ambiguity in the shower axis, which may be parametrized by $R_p$. For different values of $R_p$, the $\chi^2$ has been minimized with respect to the other parameters and is plotted in the top graph as a function of $R_p$. The same points are replotted in the lower three graphs as a function of the corresponding values for $\psi$, $X_{max}$, and energy, respectively. Note that $X_{max}$ decreases while the other parameters increase.

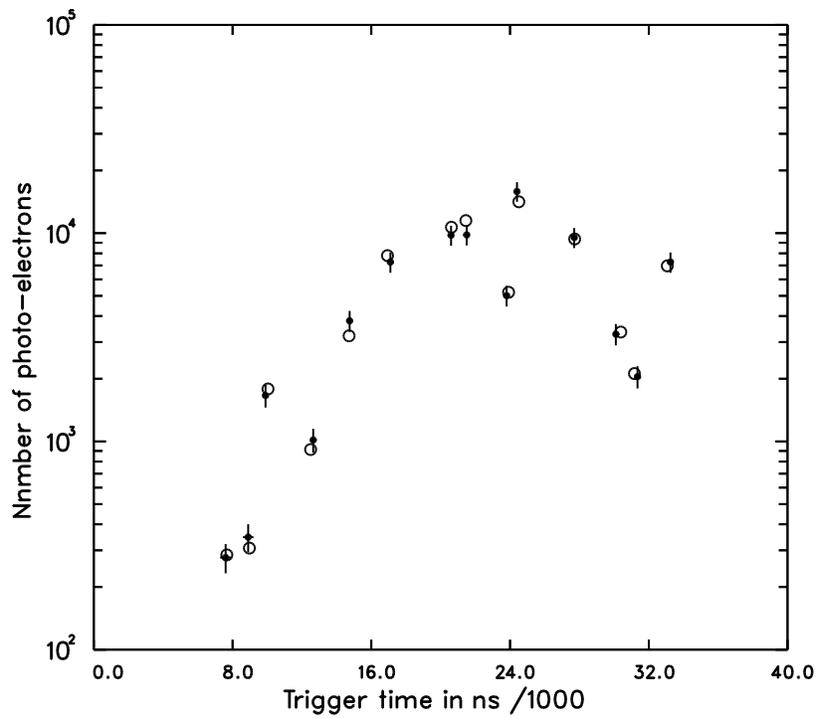

Figure 6: The data (trigger times and amplitudes for the 14 phototubes) are shown by dots with error bars in both amplitude and time (although the time error bars are smaller than the dots for some tubes). The circles denote simulated data using the best-fit shower parameters and our model of the light production, atmospheric transmission, and detector response.

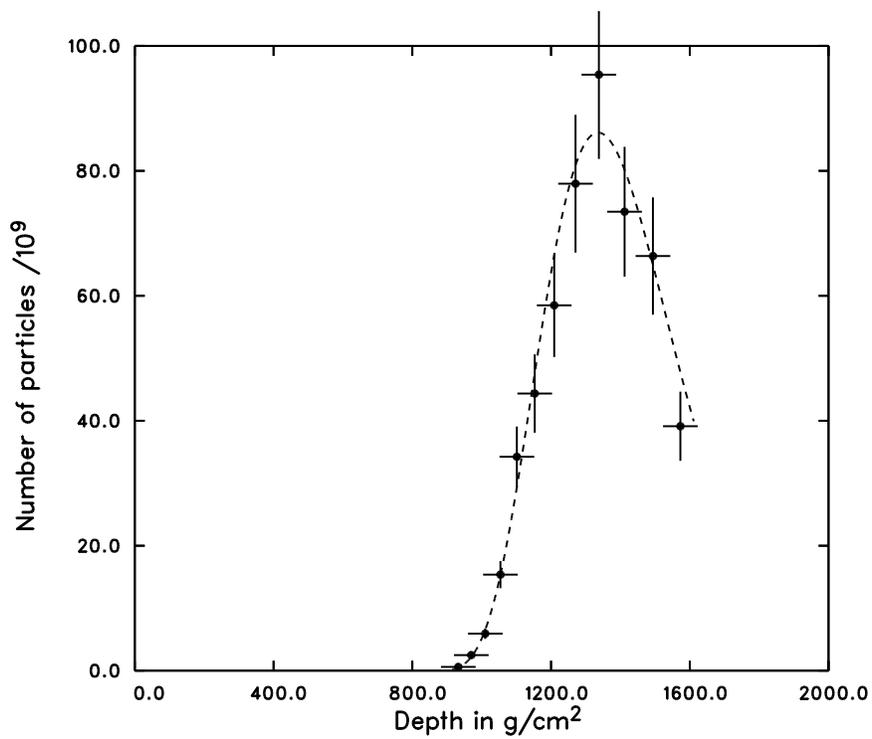

Figure 7: To account for the observed mean angular speed and the PMT amplitudes, the atmospheric depth necessarily increases if the energy is forced to lower values. Shown here is the longitudinal profile which results if the energy is forced to be only 100 EeV. The shower development starts too deep (807 gm/cm$^2$). The $\chi^2$ for this shower geometry (132) is also far from acceptable.